# GENERATING TARGET PROBABILITY SEQUENCES AND EVENTS

*Vaignana Spoorthy Ella*

*Abstract* Cryptography and simulation of systems require that events of pre-defined probability be generated. This paper presents methods to generate target probability events based on the oblivious transfer protocol and target probabilistic sequences using probability distribution functions.

**Keywords** Oblivious transfer, Probability distribution functions

## 1. Introduction

Generating target probability events has many applications in the area of system simulation, gaming, auction and key exchange (e.g. [1],[2]) as well as in cryptography, secure multi-party communication [3],[4]. The process by which nature generates probabilistic events is also of importance from a fundamental physics perspective [5]-[8] and that of neuroscience [9],[10]. Binary sequences having specific target 0's and 1's probabilities play an important role in computing, coding, communications, and cryptography [11],[12] and there is continuing search for new techniques that generate random sequences.

Here we present methods to generate target probability events based on the oblivious transfer protocol and target probabilistic sequences using probability distribution functions.

### 1.1 Oblivious transfer protocol

This protocol was introduced by Rabin [13]. In this protocol the sender sends a message to the receiver without knowing that which piece of information is being sent. The receiver receives that piece of information with certain probability. In general this probability would be ½. And the sender will never come to know whether the receiver received the message or not [14]-[17].

They are many flavors of this protocol like Rabin's protocol, "1-out-of-2 oblivious transfer", "k-out-of-n oblivious transfer", and "n-out-of -n oblivious transfer". Oblivious protocol based on key-exchange is provided in [18].

Generating probability events [7] has obvious applications in gaming as well a multiparty communications and electronic commerce. To determine if a given sequence satisfies randomness properties, one can use a variety of measures.

### 1.1.1 If we have one message secret:

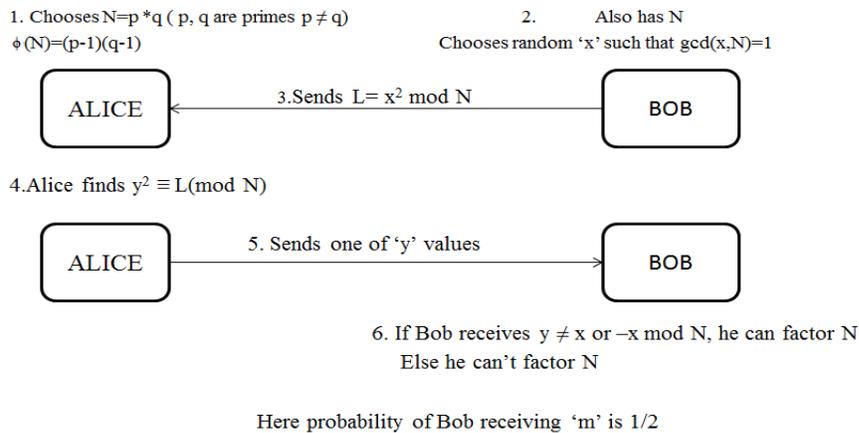

### Example 1.1.1.1

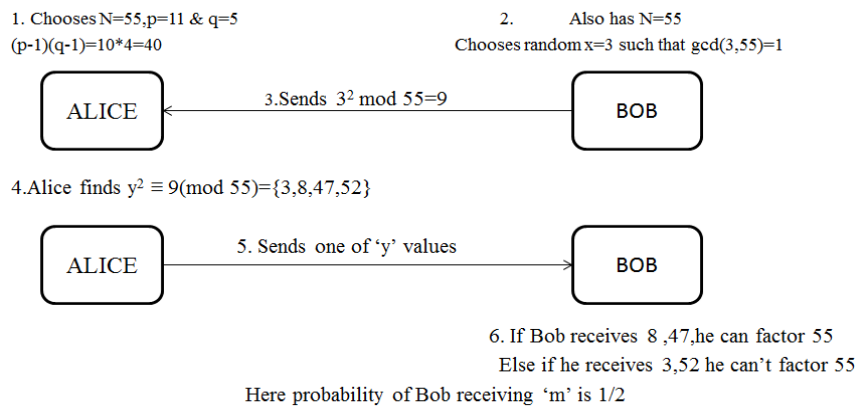

### 1.1.2 If you have two message secrets (1-out-of-2 Oblivious Transfer)

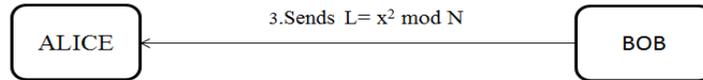

In this protocol Bob receives either '$m_0$' or '$m_1$'. However, Alice will not come to know whether Bob has received '$m_0$' or '$m_1$', but confirms either of them was received by Bob.

### 1.1.3 Generalized Oblivious Transfer

The "k-out-of-n" type of oblivious protocols falls under generalized oblivious transfer. This protocol is very useful in secure multiparty communication. In k-out-of-n protocol, the sender will be having n messages and receiver has indexed 'i' using which he selects one of the k-sized subset from 'n' messages. However, the sender will not come to know which set of k messages receiver has chosen but confirms that messages were received by the receiver.

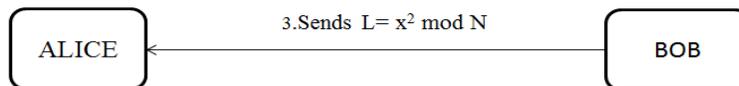
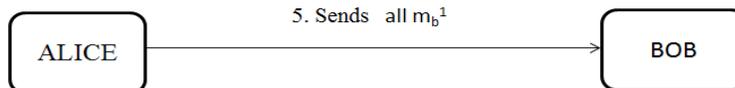

## 2. Generating Target Probabilistic Sequences

Generating a random sequence if we have given 0's and 1's sourced and destination probabilities in the sequence are shown below.

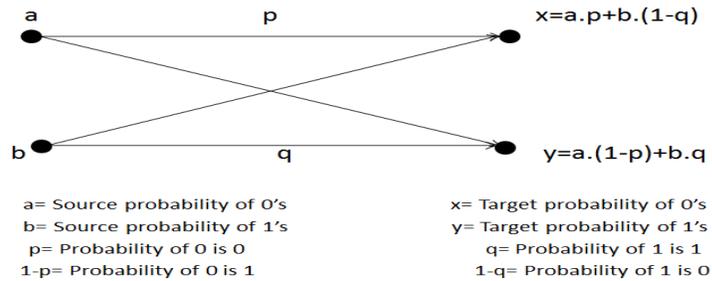

a= Source probability of 0's
b= Source probability of 1's
p= Probability of 0 is 0
1-p= Probability of 0 is 1

x= Target probability of 0's
y= Target probability of 1's
q= Probability of 1 is 1
1-q= Probability of 1 is 0

Here we know a, b, x and y. This is nothing but solving a quadratic equation and finding its solutions. We need to find p and q by solving the below equations

$$X = a \cdot p + b \cdot (1-q) \quad \text{---- (1)} \quad Y = a \cdot (1-p) + b \cdot q \quad \text{---- (2)}$$

If we solve the above equations we get many combinations of p and q values as valid solutions. Choose one of the p, q values and use them to get the output sequence with required probability as follows

Go through the input sequence considering one bit at a time.
- If the input bit is '0', toss a coin and see if the value is < p, then keep the '0' as '0' else change it to '1'.
- If the input bit is '1', toss a coin and see if the value is < q, then keep the '1' as '1' else change it to '0'.

**Example 1** Input: 0 0 0 0 0 0 0 0 0 0 0 0 0 0

- Source Probabilities:    0's: 1.0    1's : 0.0
- Target probabilities:    0's : 0.3    1's : 0.6

Error-rate calculations

Error-percentage= (Estimated value-actual value)/actual value*100

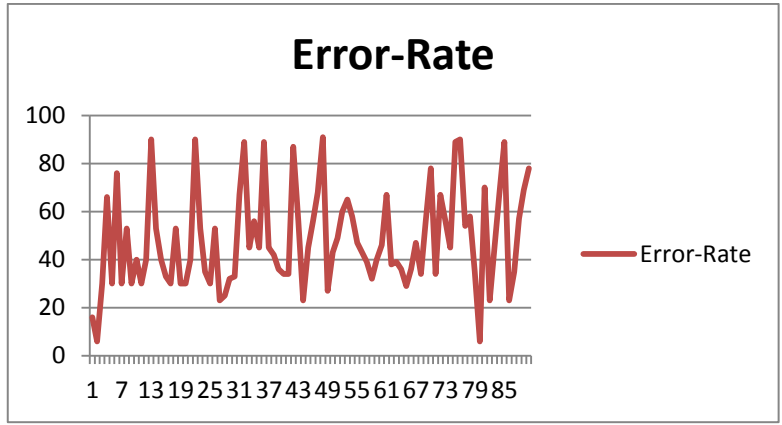

Graph 1 X-axis: iterations Y-axis: error-rate

As the length of the sequence is small, getting target probabilities in few iterations is not possible. We may or may not get the target probabilities in all iterations irrespective of p and q values.

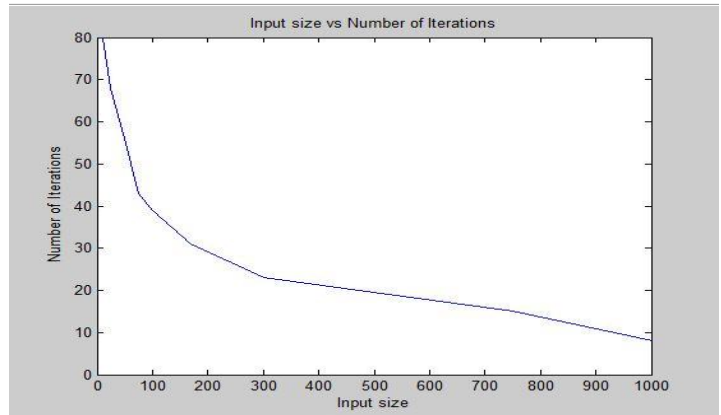

Graph 2 Input sizes Vs. Number of iterations to achieve target probability

## 3 Generating Target Probabilistic Events

Here we see how to generate different probabilistic events. In general, the power 'k' should divide ϕ (N) and probability event generated is (k-1)/k.

**Example 3.1.1**

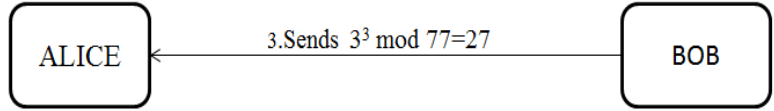

3.Sends $3^3$ mod 77=27

4.Alice finds $y^3 \equiv 27 \pmod{77} = \{3, 47, 69\}$

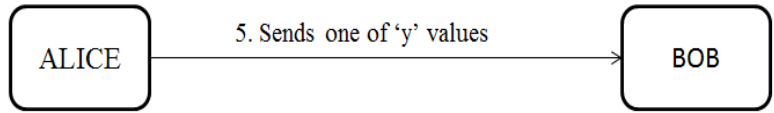

5. Sends one of 'y' values

6. If Bob receives 69, 47, he can factor 77
Else if he receives 3, he can't factor 77

Here probability of Bob receiving 'm' is 2/3

**Example 3.1.2**

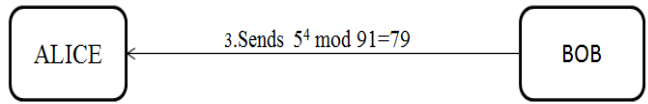

3.Sends $5^4$ mod 91=79

4.Alice finds $y^4 \equiv 79 \pmod{91} = \{5, 12, 40, 44, 47, 51, 79, 86\}$

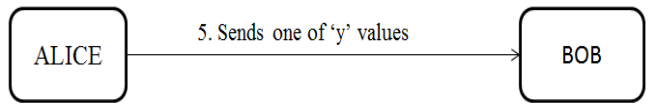

5. Sends one of 'y' values

6. If Bob receives 12, 40, 44, 47, 51, 79 he can factor 91
Else if he receives 5 or 86, he can't factor 91

Here probability of Bob receiving 'm' is 3/4

### 3.2 Dividing into Sub groups

The nine cube roots of 1 may be written as numbers and their squares like [19], [20]:

$$1, a, b, c, d, a^2, b^2, c^2, d^2 \equiv 1, a, a^2; 1, b, b^2; 1, c, c^2; 1, d, d^2$$

**Example 3.2.1** Consider 1, 9, 81; 1, 16, 74; 1, 22, 29; 1, 53, 79

Now let Bob pick one of these four sets. Then chances of Bob receiving 'm' is ¼. In general for 'r' subgroups of equation $m^k \equiv c \pmod{N}$ corresponding probability of $1/(r+1)$ [3]

## 3.3 Generalization to get target probability

### 3.3.1 Concept of serial probability Network

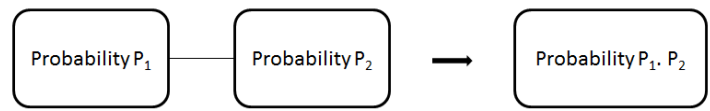

Serial Circuit probability

Probability of ½ if repeated 'n' times serially produces $1/2^n$ probability. Likewise, probability of $(k-1)/k$ if repeated 'n' times produces a final probability of $[(k-1)/k]^n$. Keeping this concept in mind we can generate the target probabilities as follows.

**Example 3.3.1.1**

If we want target probability as 3/7 we can write it as ¾ * ½. Now ¾ probability can be obtained using $m^k \equiv c \pmod{N}$ where k=4, similarly ½ probability can be obtained using k=2.

**Example 3.3.1.2**

Target probability of 3/7 can be achieved if we write 3/7 as 3/6 * 6/7. Here 3/6(=½) can be obtained for k=2 and 6/7 can be obtained for k=7

**Example 3.3.1.3**

Target probability of '1/100' can be achieved if we write 1/100 as 99/100 * 98/99 * ….½. In general, we can obtain target probability p/q using serial probabilities as shown below.

(q-1)/q * (q-2)/ (q-1) * ……till (p-i)/p. Where (p-i)/p = (k-1)/k for some integers i and k.

## 4. CONCLUSIONS

We have described methods of generating various probabilistic events that have many potential applications. These methods will be useful in solving polynomial time computable function without using any additional primitives, quantum oblivious key distribution, secure multi-party computation, and so on.

## REFERENCES

[1]   B. Schneier, Applied cryptography: Protocols, algorithms, and source code in C. John Wiley, 1996.
[2]   J. Hoffstein, J. Pipher, J.H. Silverman, An Introduction to Mathematical Cryptography. Springer, 2010.
[3]   S. Kak, Classification of random binary sequences using Walsh-Fourier analysis. IEEE Trans. on EMC, vol. EMC-13, pp. 74-77, August 1970.
[4]   S. Kak, Encryption and error-correction coding using D sequences. IEEE Transactions on Computers, vol. C-34, pp. 803-809, 1985.
[5]   R. Penrose, Shadows of the Mind. Oxford University Press, London, 1994.
[6]   S. Kak, Information, physics and computation. Foundations of Physics 26: 127-137, 1996.
[7]   S. Kak, The Architecture of Knowledge. CSC, Delhi, 2004.
[8]   S. Kak, The Nature of Physical Reality. Peter Lang, New York, 1986.
[9]   M.S. Gazzaniga, The Cognitive Neurosciences. MIT Press, Cambridge, MA, 1995.
[10]  S. Kak, Active agents, intelligence and quantum computing. Information Sciences, 128, 1-17, 2000.
[11]  S. Kak, On the method of puzzles for key distribution. Int. Journal of Comp. and Inf. Sciences, vol. 13, 1984, pp. 103-109.
[12]  S. Kak, On secret hardware, public-key cryptography. Computers and Digital Technique (Proc. IEE - Part E), vol. 133, pp. 94-96, 1986.
[13]  M.O. Rabin. Digitalized signatures and public-key functions as intractable as factorization. MIT/LCS/TR-212, MIT Laboratory for Computer Science, 1979.
[14]  M. Naor, B. Pinkas, Oblivious transfer and polynomial evaluation. ACM Symposium, 1999.
[15]  C. Bennett, G. Brassard, C. Crépeau, Practical quantum oblivious transfer. Advances in Cryptology 1992.
[16]  M. Bellare, S. Micali, Non-interactive oblivious transfer and applications. Advances in Cryptology, pages 547-557, 1990.

[17]   C. Crépeau, Equivalence between two flavours of oblivious transfers. 1998.

[18]   A. Parakh, Oblivious transfer based on key exchange. Cryptologia 32: 37-44, 2008.

[19]   S. Kak, The cubic public-key transformation. Circuits Systems Signal Processing 26: 353-359, 2007.

[20]   S. Kak, The piggy bank cryptographic trope. 2013. arXiv:1301.0313